\documentclass[10pt,conference]{IEEEtran}
\usepackage[pdftex]{graphicx}
\usepackage[usenames,dvipsnames,table,xcdraw]{xcolor}
\usepackage{array}
\usepackage{multirow}
\usepackage[normalem]{ulem}
\useunder{\uline}{\ul}{}
\usepackage{booktabs}
\usepackage{supertabular,booktabs}
\usepackage{longtable}
\usepackage{lscape}
\usepackage{cite}
\usepackage{url}
\usepackage{hyperref}
\usepackage{dirtytalk}
\usepackage{graphicx}
\usepackage{comment}
\usepackage{makecell}
\usepackage{svg}
\usepackage{amsmath}
\usepackage[caption=false]{subfig}
\graphicspath{{Images/}}
\usepackage[normalem]{ulem}
\useunder{\uline}{\ul}{}

\usepackage{tabularx}
\usepackage{balance}
\usepackage{nameref}

\begin{document}

\title{Benefits and Limitations of Remote Work to LGBTQIA+ Software Professionals}

\author{
\IEEEauthorblockN{ Ronnie de Souza Santos}
\IEEEauthorblockA{ Cape Breton University\\
    Sydney, NS, Canada \\
    ronnie\_desouza@cbu.ca }
\and
\IEEEauthorblockN{ Cleyton V. C. de Magalhães }
\IEEEauthorblockA{ CESAR School \\
  Recife, Brazil \\
  cvcm@cesar.school }
\and
\IEEEauthorblockN{ Paul Ralph }
\IEEEauthorblockA{ Dalhousie University\\
  Halifax, Canada \\
  paulralph@dal.ca } 
}






\IEEEtitleabstractindextext{%
\begin{abstract}
\textit {Background}. The mass transition to remote work amid the COVID-19 pandemic profoundly affected software professionals, who abruptly shifted into ostensibly temporary home offices. The effects of this transition on these professionals are complex,  depending on the particularities of the context and individuals. Recent studies advocate for remote structures to create opportunities for many equity-deserving groups; however, remote work can also be challenging for some individuals, such as women and individuals with disabilities. As the discussions on equity, diversity, and inclusion increase in software engineering, it is important to explore the realities and perspectives of different equity-deserving groups to develop strategies that can support them post-pandemic. 
\textit {Objective}. 
This study aims to investigate the effects of remote work on LGBTQIA+ software professionals. 
\textit {Method}. Grounded theory methodology was applied based on information collected from two main sources: a survey questionnaire with a sample of 57 LGBTQIA+ software professionals and nine follow-up interviews with individuals from this sample. This sample included professionals of different genders, ethnicities, sexual orientations, and levels of experience. Consistent with grounded theory methodology, the process of data collection and analysis was conducted iteratively using three stages of coding: line-by-line, focused, and theoretical. Member checking was used to validate the findings obtained from interpreting the experiences commented on by LGBTQIA+ software professionals.
\textit {Findings}. Our findings demonstrate that (1) remote work benefits LGBTQIA+ people by increasing security and visibility; (2) remote work harms LGBTQIA+ software professionals through isolation and invisibility; (3) the benefits outweigh the drawbacks; (4) the drawbacks can be mitigated by supportive measures developed by software companies.
\textit {Conclusion}. This paper investigated how remote work can affect LGBTQIA+ software professionals and presented a set of recommendations on how software companies can address the benefits and limitations associated with this work model. In summary, we concluded that remote work is crucial in increasing diversity and inclusion in the software industry. 
\end{abstract}

\begin{abstract}
\textit {General Abstract}. Remote work is here to stay. There is no denying it, as some software professionals would rather quit their jobs than return to the office full-time. Therefore, software companies want to understand how the remote working model can be successfully used without causing major issues. The problem is that the effects of remote work are complex because they depend on individual and group characteristics that require careful evaluation. In this scenario, one thing has been extremely positive: remote work is helping to increase diversity in software engineering by fostering new opportunities and better work conditions for individuals from equity-deserving groups, for instance, LGBTQIA+ software professionals. The software industry is overly homogeneous, most of the professionals who work in this area are heterosexual men (a reflection of the university courses on computer science and software engineering), but diversity can only be good for an area that strongly depends on creativity and innovation. What better way to innovate than putting several individuals from different backgrounds and with various experiences to work together? Remote work plays an important role in improving equity, diversity, and inclusion in the software industry. In this paper, we discuss how remote work is affecting software professionals from the LGBTQIA+ community and provide a list of recommendations to support software companies in dealing with this work model.
\end{abstract}

}

\maketitle

\IEEEdisplaynontitleabstractindextext

\IEEEpeerreviewmaketitle

\section{Introduction}\label{sec:introduction}

Equity, diversity, and inclusion (EDI) have become central topics of discussion in society. EDI is a complex phenomenon centered on developing approaches for providing equal opportunity for individuals (equity) while recognizing their personal, social, and cultural differences (diversity) and encouraging them to participate in environments and debates (inclusion)~\cite{jibc2022, suzuki2021interview}. EDI discussions are gradually increasing in computer science~\cite{silveira2019systematic}. Since technology has a direct impact on our society, the lack of diversity among software professionals can produce software products that exclude groups of users~\cite{albusays2021diversity, rodriguez2021perceived, zolduoarrati2021value}. 

Diversity is essential in developing technologies to support a pluralistic society. However, the software industry is experiencing a diversity crisis as software teams are mainly composed of heterosexual men~\cite{albusays2021diversity}. Currently, there is a lack of knowledge on how EDI should be addressed and integrated into software engineering~\cite{janzen2018reflection, zolduoarrati2021value, menezes2018diversity}. The main consensus among researchers is that diversity needs to be addressed from the academy to the software industry by increasing opportunities, improving debates, and fostering safe spaces for equity-deserving groups. 

Recently, the COVID-19 pandemic intensified discussions about EDI in software engineering. As the lockdowns forced software professionals to work from home, remote work structures created better work-life balance, more flexibility, and less commuting, which resulted in new opportunities for many underrepresented groups in the software industry: caregivers (especially mothers), LGBTQIA+ individuals (especially trans people), and people with disabilities~\cite{ralph2020pandemic, ford2021tale, albusays2021diversity}. Many software companies have permanently transitioned to remote or hybrid work, allowing software professionals to decide where they will work (e.g. in the office, from home)~\cite{barrero2021let, melin2021employees}. This uptake of remote and hybrid work may be increasing diversity and inclusion in the software industry because more job opportunities will be available for professionals that cannot afford to work in an office every day. 

Understanding the impacts of remote work arrangements on individuals from equity-deserving groups is important because equity-deserving groups experience remote work differently~\cite{ralph2020pandemic, ford2019remote}, and EDI is essential to forge a more just society. In software engineering, understanding the impacts of remote work on software professionals from equity-deserving groups is an important step toward sharpening practices, strategies, and tools required to improve EDI in the software industry. In this study, we, therefore, explore the following research question:

\smallskip
{\narrower \noindent \textit{\textbf{Research Question:} How does remote work affect LGBTQIA+ software professionals?}\par}
\smallskip

In particular, we explore the experience of software professionals from the LGBTQIA+ community because the research on EDI in software engineering that focuses on LGBTQIA+ software professionals is scarce~\cite{silveira2019systematic, rodriguez2021perceived}. From this introduction, the rest of this study is organized as follows. In Sefction \ref{sec:RelatedWork}, we discuss existing studies on EDI in software engineering. In Section \ref{sec:method}, we present the method applied in this study. In Section \ref{sec:results}, we present our findings which are discussed in Section \ref{sec:discussions}. Finally, we summarize the contributions of this study and opportunities for future research.

\section{Background} \label{sec:RelatedWork}
 
Remote work has received attention from researchers from several areas since this is expected to be the most common work arrangement in many industries~\cite{haag2021remote}. In software engineering, remote work is studied for at least two reasons. First, software professionals are more likely to continue working remotely since they are already used to this type of work structure (e.g., global software development, distributed teams)~\cite{chavez2020permanently}. Second, software professionals are responsible for creating and adapting technologies to support employees from other industries, which might also face significant post-pandemic adaptation~\cite{wang2020open}. 

Considering the first reason, as software engineering follows the world’s transformations and their impacts on society, understanding the long-term implications of the post-pandemic reality in the software industry and practitioners is crucial. Over the past two years, studies revealed that remote work could produce several benefits for software professionals. However, some limitations have also been observed in the same period~\cite{russo2021daily}. Remote work benefits software professionals by providing them with more flexibility, better work-life balance, and increased job satisfaction~\cite{ralph2020pandemic, ford2021tale}. On the other hand, remote work can foster problems at the team level, creating communication issues, increasing coordination challenges, and reducing team cohesion~\cite{desouzasantoschase, santos2022grounded}.

A significant challenge associated with studying remote work is that its effects depend on individual characteristics and the particularities of some groups of professionals~\cite{ralph2020pandemic}. For instance, remote work can be worse for women and caregivers (e.g., parents), as they might have to deal with many distractions while performing software development activities that demand high levels of concentration~\cite{ralph2020pandemic}. Remote work might create opportunities for people with disabilities~\cite{schur2020telework}; however, it can also force them to deal with poor ergonomics~\cite{ralph2020pandemic}. Remote structures give transgender software developers better control over their identities~\cite{ford2019remote}; however, in other fields, researchers are concerned that remote interactions can undermine diversity~\cite{greenwood2021sa}.

\section{Method} \label{sec:method}
Aiming to discuss the impacts of remote work on LGBTQIA+ software professionals, we conducted a qualitative study using constructivist grounded theory methodology~\cite{charmaz2014constructing}. Grounded theory methodology (GTM) is a family of methods for inductively generating theories based on the data collected from real-life contexts to identify and describe concepts, behaviors, and experiences~\cite{glaser1978theoretical}. It is widely applied in social sciences and is well-suited for exploring social, cultural, and human aspects of software development~\cite{stol2016grounded}. \textit{Constructivist} GTM adapts classic GTM to align better with a constructivist epistemology~\cite{charmaz2014constructing}; that is, the philosophical view that scientists create, rather than discover, knowledge. We use Constructivist GTM because constructivism is more consistent with our views of qualitative research, and the ostensible pragmatist philosophical underpinnings of classic GTM are by now somewhat outdated. 

GTM is appropriate for this study because it helped us understand the research problem using the experience of a small group of individuals~\cite{charmaz2014constructing}. Due to the characteristics of our targeted population, we could not expect to recruit a large number of participants. In addition, GTM effectively supported us in constructing concepts that can be transferred to other contexts through theoretical generalization~\cite{carminati2018generalizability}.

Following appropriate guidelines for GTM 
(i.e.,~\cite{charmaz2014constructing,stol2016grounded,ralph2020empirical}), we used well-defined rounds of data collection and analysis, including inductive coding, memoing, constant comparison, and theoretical sampling. This process allowed concepts to ``emerge'' from the field rather than fitting data into preconceived theories~\cite{glaser1978theoretical}.

\subsection{Participants} \label{sec:participants}
The population of interest for this study is LGBTQIA+ software professionals, including developers, QAs, analysts, designers, and managers, among others directly participating in the software development process. These professionals belong to a hidden population---a population that cannot be easily defined or enumerated based on existing knowledge---which hinders recruitment~\cite{heckathorn1997respondent}. LGBTQIA+ individuals are often treated as a hidden population in other fields because many of them are uncomfortable with discussing aspects of their sexuality due to the risk of being exposed to structural and social discrimination~\cite{hughes2021methods}. We, therefore, used a multi-pronged sampling strategy (see Section \ref{sec:dataCollection}). Nevertheless, due to the hidden and vulnerable nature of the population, we knew that recruiting a large number of participants (e.g., 1000 questionnaire respondents or 50 interviewees) would be unrealistic. 

\subsection{Instrumentation}

Our instrumentation was complicated by the desire to collect in-depth accounts of LGBTQIA+ professionals without forcing them to reveal their identities. Interviews are better for eliciting in-depth accounts; questionnaire surveys are better for anonymous participation. We, therefore, adopted both approaches to data collection, allowing participants to choose whether to fill out a mostly open-ended questionnaire survey or participate in a synchronous, semi-structured interview.

We designed the questionnaire to anonymously obtain both basic information about the participants using closed-ended questions and then elicit qualitative data about the professionals' experiences using open-ended questions. We began by asking individuals to provide their age, gender, sexual orientation, ethnicity, and country. Then, we added questions about their work and their company, including their level of education, level of experience working in the software industry, role (e.g., developer, tester, designer, manager, etc.), and the number of employees working in their company. Finally, we asked the following open-ended questions about their experience working remotely:

\begin{itemize}
    \item Would you say that remote/hybrid work benefits LGBTQIA+ individuals working in the software industry? Please, describe in what ways.
    \item Would you say that remote/hybrid work creates challenges for LGBTQIA+ individuals in the software industry? In what ways?
    \item Do you know any software professionals from the LGBTQIA+ community that benefited from  working from home on a regular basis? Tell us a bit about this.
    \item Do you know any professional from the LGBTQIA+ community that faced disadvantages caused by working from home on a regular basis? Tell us a bit about this.
    \item Is remote work more inclusive for LGBTQIA+ software professionals? In which ways?
   \item Does remote/work create more opportunities for LGBTQIA+ software professionals? How?
\end{itemize}

All questions were optional, so individuals were free to answer only the ones that they wanted to answer. At the end of the questionnaire, we asked those who would be interested in discussing more aspects of remote work and how it affects the software professionals from the LGBTQIA+ community to sign up for an interview by providing their email. The questionnaire was offered in two languages: English and Portuguese.

\subsection{Data Collection} \label{sec:dataCollection}

We used three sampling approaches~\cite{baltes2022sampling} to recruit questionnaire respondents : 

\begin{enumerate}
    \item Convenience sampling: We invited participants based on their availability. The first author is a member of the target population (i.e., an LGBTQIA+ professional who worked with software development for six years) and invited members of the population from his contact list. 
    \item Purposive sampling: We used the communication channels of a large software company in South America to advertise our questionnaire to over 1,000 software 
    professionals. This company was founded in 1996, specializes in on-demand software development, and creates technologies for clients from various sectors, including finance, telecommunication, manufacturing, and services. In addition to this company, we advertised the questionnaire in online communities from social media websites, such as LinkedIn.
    \item Snowballing sampling: We asked participants to invite other LGBTQIA+ software professionals to participate in the study. 
\end{enumerate}

We began with the questionnaire because individuals from hidden populations usually avoid outsiders~\cite{wejnert2008web}. However, members of such populations often know each other. Therefore, by applying the questionnaire first, we were able to apply sampling techniques commonly used in surveys. Once we reached the population, we could identify individuals to be interviewed, providing us with additional details beyond those collected with the open-ended questions in the survey. 

Data collection using the questionnaire happened between June 10th and July 20th, 2022. From a grounded theory method perspective, we considered the questionnaire responses as the first round of data collection. Next, we invited the participants who volunteered for interviews and conducted three rounds of interviews. These interviews were conducted online with nine participants (three per round) using Google Meet. In these three rounds, we focused on obtaining more details about the answers already provided in the questionnaire. In addition, as the rounds evolved, participants were asked about the role of diversity on software teams and the effects of remote work on software teams. The nine interviews ranged from 20 to 43 minutes. This produced four hours and 18 minutes of audio and 97 pages of transcripts.

\subsection{Data Analysis}
We started data analysis by applying descriptive statistics~\cite{george2018descriptive} to summarize the information about our sample of participants. Descriptive statistics allowed us to present the distribution of participants' answers regarding their personal and professional profiles. 

Following this, we analyzed the answers to open-ended questions in the questionnaire (first round). Then, we continued using qualitative data analysis in the interviews (following three rounds). Qualitative analysis was conducted inn three stages (following~\cite{charmaz2014constructing}): line-by-line coding was focused on identifying initial codes and building concepts within our data; focused coding supported the establishment of connection among the concepts and the definition of high-level categories of concepts; theoretical coding was used to emphasize the core category and explain the main story arising from the data.

In this process, we began with line-by-line open coding, which allowed us to identify and refine emerging concepts from the answers to the open-ended questions in the survey. We repeated this strategy in the first round of interviews through a systematic process supported by audio recordings, which helped identify the properties of codes and their meanings. Following the second round of interviews, we began categorizing these codes and establishing connections using focused coding. Focused coding was performed through repeated analyses and comparisons among emerging categories. This process continued during the third and final round of interviews. When these transcripts were integrated into the analysis, we used theoretical coding to rearrange the categories and highlight a central story about the experience of our participants.

\subsection{Ethics}
The questionnaire began by presenting the research goal and the study's relevance to the software industry. Respondents needed to agree explicitly to provide data to the study by selecting the option at the beginning of the questionnaire. Later, at the end of the questionnaire, participants could volunteer to be interviewed by providing their email addresses. At the beginning of each interview, the interviewer explained the goal of the research one more time and the need to collect more data in addition to the questionnaire before interviewees again orally consented to participate. Participants were guaranteed data confidentiality and de-identification of their quotes. They were also informed about the voluntary nature of their participation and the right to stop the interview and withdraw from the research at any time. No participants withdrew. The third author's university research ethics board approved this research.

\section{Findings} \label{sec:results}
This section presents the demographics of our sample, followed by the categories that emerged from the analysis; that is, the benefits and limitations of remote work for LGBTQIA+ software professionals. Our findings are illustrated with quotations extracted from questionnaire responses and interviews and presented in Tables \ref{tab:benefits} and \ref{tab:limitations} below. Some quotations have been translated into English and may read awkwardly as we have transcribed and translated them as accurately as possible.

\subsection{Demographics}
Our survey reached 81 individuals. However, some participants were excluded from the survey for not being part of the target population:

\begin{itemize}
\item one woman and one man identified themselves as heterosexual, despite stating  that they were software professionals from the LGBTQIA+ community in the screening questions. 
\item 22 individuals only completed the screening questions; they answered neither demographic nor open-ended questions. 
\end{itemize}

However, our sample of 57 individuals varies as not all participants answered the entire questionnaire. For instance:

\begin{itemize}
\item 57 (100\%) individuals provided their level of education.
\item 56 (98\%) individuals provided their role in the software team.
\item 53 (93\%) individuals provided their age, gender, and ethnicity.
\item  52 (91\%) individuals provided their country and their level of experience. 
\item 45 (79\%) individuals provided information about the number of employees in their companies.
\item 25 (44\%) individuals answered the question about their sexual orientation.
\item 13 (23\%) individuals answered at least one question about how remote work affects LGBTQIA+ software professionals.
\item 10 (17\%) individuals provided us with an email for an interview.
\end{itemize}

Participants' ages varied from 21 to 50 years old. Participants had an average of 5.7 years of work experience in software development; the most experienced professional in the sample has worked in the software industry for 25 years, and the least for one year. Table \ref{tab:benefits} Table Tables \ref{tab:Demographics} presents more details about our sample and its variations. Finally, we were able to interview 16\% of the survey (9/57), who provided more details on the topics addressed in the research. 

\begin{table}
\centering
\caption{Demographics}
\renewcommand{\arraystretch}{1}
\label{tab:Demographics}
\begin{tabular}{llr}
\toprule
\multirow{4}{*}{ \textbf{Gender} } 
& Men & 34\\
& Women & 14\\ 
& Non-binary & 3\\ 
& Not informed & 4\\ \midrule 
\multirow{6}{*}{ \textbf{Ethnicity} } 
& White & 34 \\
& Mixed ethnic  & 13 \\
& Black & 2 \\
& Arab & 2 \\
& Asian & 1 \\
& Latino & 1 \\
& Not informed & 4 \\\midrule 
\multirow{6}{*}{ \textbf{Sexual Orientation} }
& Gay & 13 \\
& Bisexual & 7 \\
& Lesbian & 3 \\
& Pansexual & 1 \\
& Asexual & 1 \\ 
& Not informed & 32 \\ \midrule 
\multirow{6}{*}{ \textbf{Education} }
& High-School & 8 \\
& Bachelor  & 30 \\
& Post-baccalaureate & 12 \\
& Master & 6 \\
& Ph.D. & 1 \\ \midrule 
\multirow{7}{*}{ \textbf{Roles and Activities$^a$} } 
& Testing & 28 \\
& Programming  & 25 \\
& Requirements & 17 \\
& Design & 14 \\
& Architecture & 10 \\
& Management & 3 \\
& Not informed & 1 \\\midrule 
\multirow{4}{*}{ \textbf{Location} } 
& Brazil & 44 \\
& Canada & 2 \\
& Portugal & 2 \\
& Sweden & 2 \\
& Netherlands & 1 \\
& US & 1 \\
& Not informed & 5 \\\midrule
\multirow{5}{*}{ \textbf{Company Size} } 
& 0 - 9 employees & 1 \\
& 10 - 99 employees & 3 \\
& 100 - 999 employees & 5 \\
& 1,000 - 9,999 employees & 35 \\
& More than 10,000 employees & 1 \\
& Not informed & 12 \\
\bottomrule
\end{tabular}

\flushleft
\footnotesize{Notes: $^a$some professionals reported performing multiple activities or having more than one roles in the team}
\end{table}


\begin{table*}[t!]
\caption{Benefits of Remote Work for LGBTQIA+ Software Professionals}
\label{tab:benefits}
\renewcommand{\arraystretch}{1.3}
\begin{tabularx}{\textwidth}{>{\raggedright\arraybackslash}p{1.8cm}>{\raggedright\arraybackslash}p{3cm}X}
\toprule
\textbf{Construct} & \textbf{Definition} & \textbf{Example Evidence} \\

\midrule
\multirow{5}{1.8cm} {Job Opportunities} & \multirow{1}{3cm}
{The availability of employment that LGBTQIA+ software professionals can easily access.}
& ``it gives professionals the chance to work in more inclusive companies that are distant from their cities.''~(P10) \\
&& ``my girlfriend is from another state, and now I am living with her and working remotely.''~(P15) \\
&& ``software companies are in need of talented professionals and this [remote work] can be a good opportunity for people from the LGBTQIA+ community.''~(P16) \\

\midrule
\multirow{2}{1.8cm} {Engagement} & \multirow{1}{3cm}
{The level of interaction of LGBTQIA+ software professionals with their teammates.}
& ``it increased the engagement of those who were afraid of participating before [in-person]''~(P12) \\
&& ``remote work allowed me to engage with new people, cultures, experiences and different realities [in the team].''~(P16) \\

\midrule
\multirow{5}{1.8cm} {Identity Disclosure Control} & \multirow{5}{3cm}
{The individual decision of concealing or communicating and expressing their gender and sexual orientation.} 
& ``virtual tools allow people to communicate without having to expose themselves either for fear or as an option.''~(P12)\\
&& ``I can choose how I show up on camera, to exclude gender identifiers that make me dysphoric or communicate a gender that is incorrect.''~(P07)\\
&& ``relying on slack has allowed our tabs/NB colleagues to state their pronouns in their bio for easy reference.''~(P11)\\

\midrule
\multirow{2}{1.8cm} {Physical Safety} & \multirow{1}{3cm}
{The individual perception of being safe and protected from any risk to their physical integrity.} 
& ``we don’t have to be exposed to public transportation and streets where trans people are killed as if this was a sport.''~(P03) \\ 
&& ``For those who don’t belong in the heteronormative context, working from home is safer.''~(P07) \\  

\midrule
\multirow{5}{1.8cm}{Toxicity Avoidance} & \multirow{5}{3cm}
{The sentiment of being distant from other people's negative comments, attitudes, and actions.} 
& ``Small types of violence that can happen in-person [such as comments and staring] are less common in virtual environments on Slack or email.''~(P03) \\
& & ``Remote or hybrid work reduces situations where people keep staring at you or making you feel uncomfortable.''~(P04) \\
& & ``In my experience, it reduces the chance of me suffering discrimination because everything is recorded (videos, screenshots, etc.)``~(P10) \\
\bottomrule
\end{tabularx}
\flushleft
\end{table*}

\subsection{Benefits of Remote Work to LGBTQIA+ Software Professionals}
Participants pointed out benefits that were observed in the general professional context (e.g., the increase in job opportunities for the community) as well as benefits observed at the individual and team levels.
Table II summarizes the benefits identified in this study and the evidence obtained from participants' experiences. These benefits are Job Opportunities, Engagement, Identity Disclosure Control, Physical Safety, and Toxicity Avoidance. All of these benefits are extremely valuable for LGBTQIA+ software professionals because the software industry is still an area dominated by heterosexual men. 

Remote work increases access to job opportunities in the software industry, and these benefits LGBTQIA+ individuals greatly. In general, this could be seen as a simple correlation; that is, more remote jobs available means more jobs available for LGBTQIA+ professionals. However, for LGBTQIA+ individuals, this aspect goes beyond the increase in the number of positions. These are professionals that have suffered discrimination for years. Our participants explained that online recruitment tends to be slightly less biased by physical characteristics. Moreover, LGBTQIA+ professionals can apply for jobs previously available to those living in major urban centers. This means that they can choose to work where they always lived instead of adapting to new places or having to move away to unfriendly locales.

Meanwhile, for many LGBTQIA+ individuals, disclosing their identity is never easy, especially when entering a new environment. Trans people and non-binary are apprehensive about preconceived judgments about their physical appearances. Gender-affirming care (e.g., medical treatments and therapy) is an additional concern for software professionals who are transgender. Other individuals from the community who do not fit into the gender binary (e.g., male or female stereotype) suffer from fear and uncertainty in relation to how they will be seen and treated by co-workers. Remote environments established a place where LGBTQIA+ software professionals feel comfortable as they can control their identity disclosure, for instance, by deciding when they want to use their cameras. 

Furthermore, the remote work structure allows LGBTQIA+ software professionals to regulate the level of interaction and contact with their teammates to the point that they feel comfortable around them. Our participants revealed that their engagement with other team members increased while working remotely because they could gradually get used to their co-workers and engage accordingly. Joining a software team before (i.e., in person) was challenging for LGBTQIA+ professionals because they were worried about acceptance in the software industry work environment since it is essentially heterosexual. In contrast, in remote environments, they will first become comfortable with their team, then engage, and once they feel included by the team, they can finally interact in-person. This is a \textit{know me before you judge me} effect that cannot be afforded in primarily in-person environments.

\begin{table*}[ht!]
\caption{Limitations of Remote Work for LGBTQIA+ Software Professionals}
\label{tab:limitations}
\renewcommand{\arraystretch}{1.3}
\begin{tabularx}{\textwidth}{p{2.0cm}p{4cm}X}
\toprule
\textbf{Construct} & \textbf{Definition} & \textbf{Example Evidence} \\
\midrule
\multirow{3}{1.8cm} {Self-isolation} & \multirow{2}{3cm}
{The strategy of increasing isolation seeking to feel safe from external attacks.}
& ``it is not all positive as it allows us [by option] to reduce our contact with people from other ethnicities, sexual orientation, social classes, therefore, reducing our adaptation  capability.''~(P01) \\
&&``if you are working from home because in-person you need to hide any of your characteristics, then, I don't see it as inclusion.''~(P04) \\

\midrule
\multirow{3}{1.8cm} {Invisibility Feeling} & \multirow{2}{3cm}
{The impression of being alone, with little or no contact with other members of the LGBTQIA+ community in the company.}
& ``it reduces our ability to adapt (…) and experience beyond the ‘bubble’ that we live in.''~(P01) \\
&& ``sometimes it feels very lonely and difficult to be just around a family that is not very supportive.''~(P08)  \\
&& ``workplaces to talk about these themes are scarce and talking about it in remote environments is even rarer.''~(P12)\\
\bottomrule
\end{tabularx}
\flushleft
\end{table*}

Many LGBTQIA+ individuals, including those working in the software industry, still feel unsafe and afraid of physical attacks. Transgender professionals who participated in this study reported the risks that they are exposed to every day, in particular, those who live in, according to them, the country where \textit{more trans women are killed in the world}. Working from home reduces the exposure of these professionals to unsafe environments, such as public transportation during peak hours in countries that are more dangerous to them. By not having to go to the office daily, these professionals can use safer alternative commuting, such as taxis and private rides, whenever they need to be in the office.

Office environments can be extremely uncomfortable for LGBTQIA+ professionals who are exposed to coworkers' negative comments, attitudes, and actions. These toxic behaviors vary from the act of being constantly observed to being a target for constant commentaries in the office. Although this is not an act of physical violence, such conduct affects individuals' well-being and their sense of belonging in relation to the company. Hybrid work allows LGBTQIA+ software professionals to decide the best moment to be in the office and avoid toxicity from others when necessary. 

The above-cited benefits are extremely valuable for LGBTQIA+ software professionals because the software industry is still an area dominated by heterosexual males, which ends up undermining the inclusion of other groups of individuals, even indirectly. Remote work and its benefits, therefore, create more opportunities for diversity in software engineering while also helping software teams to be more inclusive.

\subsection{Limitations of Remote Work to LGBTQIA+ Software Professionals}
Although remote work produced benefits that improve the experience of LGBTQIA+ individuals working in the software industry, there are also observable limitations reported by our participants that cannot be overlooked. Most of the benefits described above are related to LGBTQIA+ software professionals distancing themselves from their coworkers and organizations for their protection. This aspect creates limitations that need to be managed at the team and organizational levels, namely, feelings of invisibility and self-isolation. 
 
By self-isolating, individuals might shield themselves from unsafe and uncomfortable situations. However, one participant reported this could be experienced as a \textit{fake protection} because the elements of violence against LGBTQIA+ software professionals will not disappear; they would just be obscured. Therefore, it is essential that software companies design strategies to increase inclusion so that professionals will feel welcome in the workspace; such that remote work is a choice, not an escape.
 
In addition, not all LGBTQIA+ software professionals have supportive families. Some of them encounter, in their companies and their teams, people that they can rely on. Remote structures create barriers for LGBTQIA+ software developers to meet with other members of the community that also work in software development, which creates the feeling of invisibility. Again, individuals reported that software companies and not themselves should be responsible for designing strategies to increase the visibility of this group. 

\subsection{How does remote work affect LGBTQIA+ software professionals?}
Working from home affects LGBTQIA+ software professionals in different ways. Primarily, our data reveals these professionals can benefit greatly from the flexibility provided by this work structure, and the opportunity of choosing where to work supports these individuals in dealing with several struggles faced for years by the community in general. However, remote work also can be associated with advantages that can be observed at the individual level and that require organizational actions to smooth such problems. 

In general, remote work increased the access of LGBTQIA+ software professionals to work opportunities. In particular, those who live in suburban areas and had difficulty joining the software industry before are now finding jobs in software companies without leaving the safety of their communities. In addition, remote work has allowed LGBTQIA+ software professionals to avoid acts of violence, both physical and emotional, resulting from discrimination in the workspace and the commuting between home and work. These are advantages of remote work that organizations might explore to develop strategies that will increase diversity in the software industry by bringing talented LGBTQIA+ professionals to work in the area. 

Merely facilitating LGBTQIA+ professionals' access to software industry jobs is insufficient to guarantee fairness in an environment that has been acutely homogenous and unfriendly to equity-deserving groups. Considering this aspect, remote work turned out to be more inclusive as it facilitates LGBTQIA+ individuals' control over their identities and regulation of their interactions with their teams based on how accepted they feel. However, inclusion in the software industry depends on the organizations' attitudes to create strategies that reinforce the importance of embracing diversity in the workspace. Such strategies should prioritize communication and networking, allowing LGBTQIA+ software professionals to develop connections within the organization, therefore, avoiding isolation. 

Two out of the three EDI principles were observed in this research, namely, diversity and inclusion. Although no aspects of equity were revealed in this study, our findings suggest that diversity and inclusion are critical elements to increasing LGBTQIA+ visibility in the software industry, which is fundamental for ensuring fair treatment and opportunities despite professionals' gender and sexuality. Visibility is an essential aspect of reducing intolerance, unfairness, and inequity against LGBTQIA+ people, especially in environments predominantly heterosexual, such as the software industry. Remote work plays a vital role in this matter. 

\section{Discussion} \label{sec:discussions}
\begin{figure*}[t]
\centering
\centerline{\includegraphics[width=1.0\textwidth]{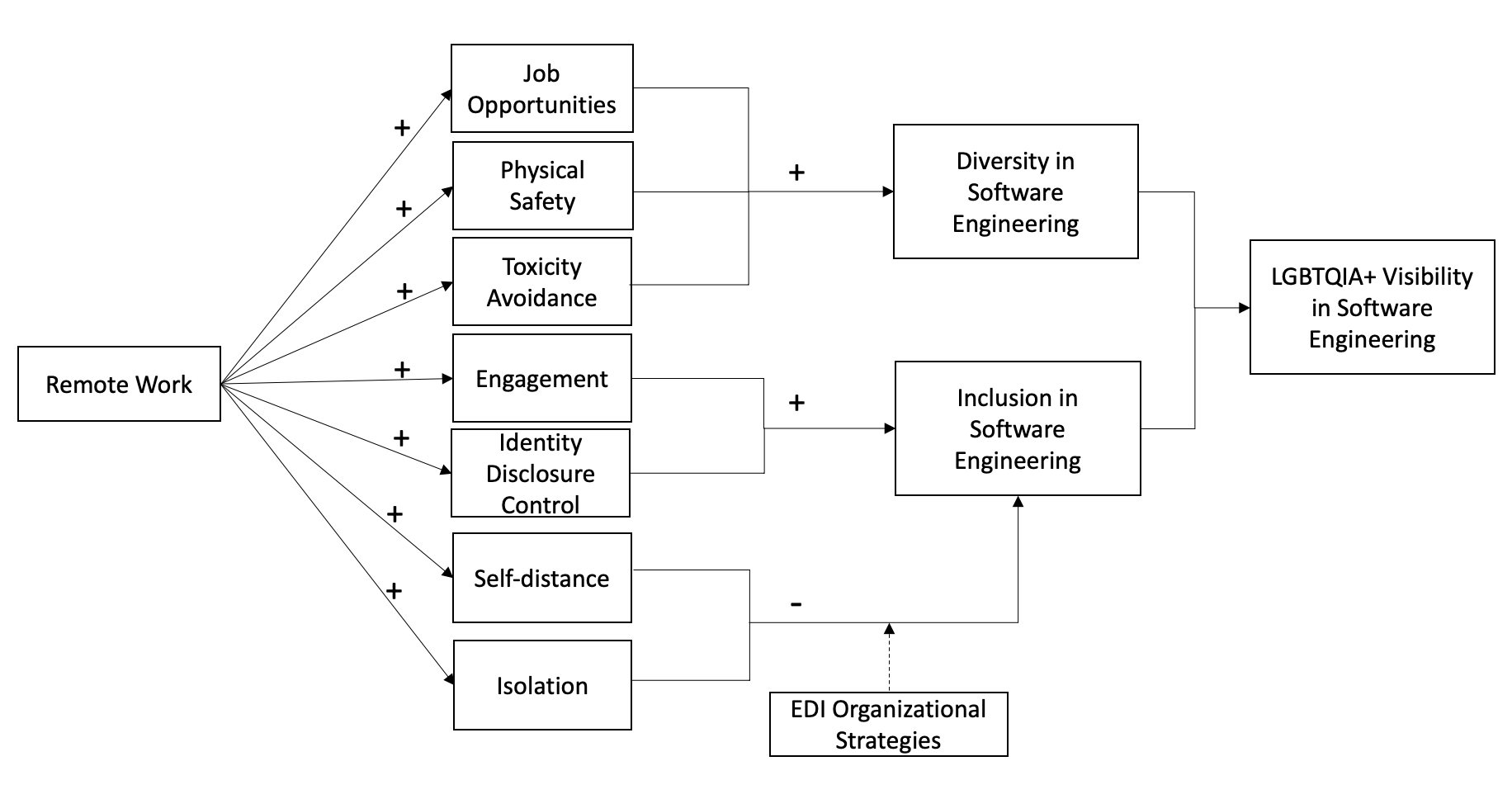}}
\caption{The effects of remote work on LGBTQIA+ software professionals}
\label{fig:theory}
\end{figure*}

Our investigation demonstrated that remote work could benefit LGBTQIA+ software professionals greatly. In general, remote work supports a structure that improves LGBTQIA+ visibility in the software industry through increasing diversity and inclusion, supported by the benefits observed in this study as summarized in Figure \ref{fig:theory}. and discussed below.   

\subsection{Re-intergrating into Prior Literature}
Diversity and inclusion are crucial factors for organizations to be more competitive~\cite{ferdman2014diversity}. Diversity is a core element for developing new ideas, which is the key to innovation~\cite{singh2018lgbt}, while inclusion supports productivity, talent retention, and engagement~\cite{schawbel2012companies}.

Diversity is fundamental for innovative environments as innovation is the enabler of business transformation in an ever-changing world where technologies (e.g., products, processes, or services) constantly evolve~\cite{ciriello2018digital}. Diverse teams improve innovation because they are more creative, more effective, and better coordinated~\cite{rodriguez2021perceived, karen2022retaining}. Remote work creates opportunities for software teams to be more diverse by adding the experiences of LGBTQIA+ professionals. LGBTQIA+ individuals are reported to be more creative and cope with high levels of autonomy, which are two essential aspects of software development nowadays, in particular in agile environments~\cite{ross2012imagined}.

Remote work supports the inclusion of LGBTQIA+ professionals in software teams, as these professionals can slowly adjust to their team (e.g., controlling the camera, engaging via chat or call) to the point that they feel comfortable enough to interact regularly, different from in-person team interaction, which allows no adjustments. Inclusion is important to software companies because it enhances organizational commitment, consequently increasing productivity, job satisfaction, and retaining talented professionals in software teams. Although in this study, we are explicitly discussing inclusion from the perspective of LGBTQIA+ software professionals, perceived inclusion is a factor that can affect all individuals in an organization~\cite{chen2018does}. 

As remote work fosters diversity and inclusion of LGBTQIA+ software professionals, we expect that LGBTQIA+ visibility will grow gradually in the software industry. On the one hand, visibility is essential in strengthening the sense of individual belonging and security for LGBTQ+ people, therefore improving several aspects of software development, such as teamwork and team resilience~\cite{sinton2021increasing}. On the other hand, this visibility is expected to transcend the boundaries of software development environments and become more frequent in the software products and technologies that nowadays impact several aspects of our society (e.g., work, education, and leisure~\cite{albusays2021diversity}), producing diverse and inclusive solutions achieving individuals from a variety of profiles. 

\subsection{Recommendations for Industry}

Our findings demonstrate several aspects of the work the software organizations can conduct to improve the experience of LGBTQIA+ software professionals. Software companies can benefit significantly from using remote work to compose highly diverse teams that reflect our multifaceted society. Our findings also highlight the importance of actions to improve inclusion in the software industry.

Based on our findings and regarding LGBTQIA+ software professionals, we recommend that software companies:

\begin{itemize}
\item Apply unbiased recruitment and hiring process to avoid discrimination.
\item Create inclusive onboarding processes, including EDI training for software teams.
\item Develop democratic remote work structures, allowing professionals to choose their workspace, which will help LGBTQIA+ software professionals to better deal with violence, toxicity, and other problems related to in-person work. 
\item Understand the specific needs of LGBTQIA+ professionals and provide them with the appropriate support, e.g., regarding cameras and video calls, and gender-affirming care, among others.  
\item Foster a culture of diversity and inclusion that embraces and welcomes LGBTQIA+ professionals. 
\item Support the creation of channels and committees to help LGBTQIA+ professionals within the company to connect among themselves and others, thus, avoiding isolation.
\item Celebrate diversity and inclusion, improving the visibility of LGBTQIA+ professionals in the company. 
\item Acknowledge the role of diversity in developing innovative technologies in modern society, thus, increasing the interest of equity-deserving groups in software engineering, e.g., students.
\end{itemize}

We understand that most of these recommendations only apply to organizations that are already LGBTQIA+ friendly and many software companies still discriminate against LGBTQIA+ individuals. We expect that the findings presented in this research, in particular, the discussions on how diversity is essential to the development of software for our modern society help policymakers to build strategies to change this reality.

\subsection{Implications for Research}

Our findings also have implications for researchers. Studies focused on LGBTQIA+ software professionals are scarce. We contribute to the theme with a comprehensive investigation focused on the effects of remote work on these professionals by exploring the experience of several groups of individuals that compose this community. To the best of our knowledge, this is the first study to have such a diverse population of LGBTQIA+ individuals in software engineering since we were able to reach lesbians, gays, bisexuals, transgenders, asexuals, and pansexuals, among others. 

Our study highlights the need for more studies of equity in software engineering for LGBTQIA+ professionals. Recently, studies have been addressing equity in the context of others underrepresented groups. We highlight the importance of increasing research efforts on addressing equity to reach the LGBTQIA+ community, while our findings are expected to initiate further discussion on this theme in academia.    

As demonstrated in Fig. \ref{fig:theory}, there is a complex causal web relating remote work to outcomes that benefit LGBTQIA+ software professionals. We focused on one outcome: visibility. There are many other outcomes, and the structure of this causal web needs more investigation and discussion. Furthermore, a qualitative survey is a suitable strategy to formulate this type of theory; however, not suitable to test it. Testing requires different types of research, including quantitative approaches, which, again, are sorely needed. 

Finally, our study is an example of how to deal with hidden populations in software engineering, in particular when addressing a group of individuals who are sensitive in relation to data collection strategies. Our data collection strategy can be used to guide other studies that deal with complex hidden populations. However, we recommend that software engineering researchers explore the well-established literature on hidden populations.

\subsection{Limitations}

Regarding the quality aspects of our method and to avoid threats to the study's validity in terms of credibility, originality, resonance, and usefulness, we followed Charmaz’s~\cite{charmaz2014constructing} criteria to evaluate grounded theory studies. To support the quality and credibility of our findings, we provided direct quotations from questionnaires and interviews to illustrate the interpretation of our participants' experiences. This interpretation was accessed by conducting member-checking with three interviewees who agreed to participate by commenting about the obtained categories after a brief presentation and explanation of the findings. 

The main limitation of this study is the number of participants from whom we were able to collect data. Investigating LGBTQIA+ software professionals is challenging because many members of this group avoid exposure in predominantly heterosexual environments like most software engineering companies. We tried to mitigate this challenge by collecting data both anonymously using a questionnaire and subsequently interviewing those who provided their contact for further discussions. This allowed us to recruit 57 individuals. Obviously, no 57 people could fully represent the experiences of a global, internally diverse group. 

Furthermore, 77\% of our participants were from Brazil, and  much of their experience depends on their local context. Although software teams worldwide often apply similar processes, and even though most of our participants have international clients, which brings them close to other cultures, several aspects of equity, diversity, and inclusion depend on organizational behaviors that are more influenced by regional factors than software development processes. Therefore, transferring our results to other regions may present challenges. We tried to mitigate this limitation by offering the questionnaire in English and advertising across social media and international forums of software professionals. In the end, however, most participants were from Brazil.

That said, representative sampling is simply not a goal of constructivist research, which instead focuses on generating transferable (not statistically generalizable) concepts and theories. Here, we acknowledge that researchers may face challenges transferring our results to different cultural regions.

\subsection{Future Work}
Equity, diversity, and inclusion in software engineering are topics with many opportunities for investigation. There are many gaps in this theme in the software industry. Additionally, it also requires investigations in the academic context, especially on how to increase diversity and inclusion in software engineering courses. 

Furthermore, our findings suggest several immediate research directions:

\begin{itemize}
\item Conduct an in-depth investigation of the factors revealed in this study and their effects on several aspects of software development; for example, the relationship between 
diversity, equity, and inclusiveness on the one hand and team conflicts, team resilience, and software practices on the other.

\item Examine the generalizability of the proposed theory using quantitative studies such as a global questionnaire survey.

\item Investigate the perspective of heterosexual professionals about EDI and its impacts on software development since people who are not part of the LGBTQIA+ community demonstrated interest in participating in this study.

\item Investigate strategies to improve equity in the software industry aiming to increase the access of LGBTQIA+ professionals to job opportunities in the area. 
\end{itemize}

Finally, we are interested in exploring EDI strategies that are being successfully applied in other industries to determine how they can be transferred to software companies by applying transformative research methods (e.g., action research ~\cite{wohlin2021guiding}).

\section{Conclusion} \label{sec:conclusion}
We investigated the effects of remote work structures on LGBTQIA+ software professionals. Using grounded theory methodology, we explored the experience of 57 individuals from different groups within the community, e.g., lesbians, gays, bisexuals, and asexuals, including both cisgender and transgender people. 

We found that remote work has a crucial role in increasing diversity and inclusion in the software industry. Remote work improves LGBTQIA+ software professionals' job opportunities and controls over their identities and interaction with other professionals. However, remote work may also create barriers for LGBTQIA+ software professionals by increasing isolation and self-distancing. Software companies need to develop practices to increase LGBTQIA+ visibility among employees. Nevertheless, the benefits of remote work appear to outweigh its drawbacks for most of our participants. Our analysis also demonstrates that software development can benefit significantly from more diverse teams, including improving aspects related to innovation, problem-solving, teamwork, and team resilience.

\section*{Data Availability} \label{sec:DataAvailability}
Supplementary material is available on Figshare, including forms, survey questionnaire, and de-identified data collected from professionals: \url{https://figshare.com/projects/Benefits_and_Limitations_of_Remote_Work_to_LGBTQIA_Software_Professionals/157305} 

\section*{Acknowledgments}
This project was supported by NSERC Discovery Grant RGPIN-2020-05001 and Discovery Accelerator Supplement RGPAS-2020-00081. The authors would like to thank all of the people who participated in this study, and for all LGBTQIA+ software professionals out there, we would like to say that we see you. You are not alone!

\ifCLASSOPTIONcaptionsoff
  \newpage
\fi

\balance
\bibliographystyle{IEEEtran}
\bibliography{bib.bib}

\end{document}